# Towards ALMA2040

## An update from the European community and invitation to contribute

**Stefano Facchini** (U. Milan), **Jacqueline Hodge** (Leiden U.), **Jes Jørgensen** (U. Copenhagen), **Eva Schinnerer** (MPIA), **Gie Han Tan** (TU Eindhoven), **Tom Bakx** (Chalmers), **Andrey Baryshev** (U. Groningen), **Maite Beltran** (INAF Firenze), **Leindert Boogaard** (Leiden U.), **Roberto Decarli** (INAF Bologna), **María Díaz Trigo** (ESO), **Jan Forbrich** (U. Hertfordshire), **Peter Huggard** (RAL), **Elizabeth Humphreys** (ESO), **Violette Impellizeri** (ASTRON), **Karri Koljonen** (NTNU), **Kuo Liu** (MPIfR), **Luca Matrà** (Trinity College), **Miguel Pereira Santaella** (IFF-CSIC), **Arianna Piccialli** (BIRA-IASB), **Gergö Popping** (ESO), **Miguel Querejeta** (OAN), **Miriam Rengel** (MPS), **Francesca Rizzo** (U. Groningen), **Lucie Rowland** (Leiden U.), **Hannah Stacey** (ESO), **Wouter Vlemmings** (Chalmers), **Catherine Walsh** (U. Leeds), **Sven Wedemeyer** (U. Oslo), **Martina Wiedner** (Obs De Paris)

## Abstract

Over the last 15 years, the Atacama Large Millimeter/submillimeter Array (ALMA) has revolutionized astrophysics by providing unprecedented resolution and sensitivity in observing the cold universe, including the formation of stars, planets, and galaxies. With groundbreaking discoveries ranging from the first detailed images of protoplanetary disks to the kinematics of galaxies in the Epoch of Reionization, ALMA has showcased the vast discovery potential of the (sub-)mm wavelength regime. However, in another 15 years from now—in the 2040s—the science landscape will have changed dramatically as new major observational facilities will have started their operations or have come towards advanced maturity in their scientific outcome (e.g., JWST, Rubin Observatory, ELT, Euclid, Gaia, Plato, Ariel, Roman Space Telescope, SPHEREx, LiteBIRD, LISA, SKA and others). At the same time, ALMA's current Wideband Sensitivity Upgrade will have been in place for ~10 years, and ALMA itself will have been operational for 30 years. To fully exploit this era, the community needs a next-generation facility operating at (sub-)mm wavelengths with capabilities far beyond those possible within ALMA's current infrastructure.

To this end, ALMA2040 [1] is a community-driven initiative to **define the key scientific questions of the 2040s** and **translate them into a technical vision for a next-generation transformational (sub-)millimeter facility**. Our goal with this document is to summarize the current status of the effort, synthesize outcomes from the 2025 workshops, outline next steps toward a reference design concept, and invite broad participation from the global mm/sub-mm community to help shape this future facility.

In the following we provide details on the process and scope. We invite everyone who wishes to join the effort and/or contribute to the dedicated White Papers planned for 2026.

# 1. About ALMA2040

The ALMA2040 initiative began in early 2025 within the European (sub-)millimeter community in response to ESO's Expanding Horizons initiative [2] (anticipated deadline: 1 June 2027). The ALMA2040 effort is coordinated through a Steering Committee and supported by nine Scientific Working Groups and four Technical Working Groups, with guidance from an Advisory Board of external experts.

The Scientific Working Groups (SWGs) represent the scientific backbone of the initiative. Established in early 2025, the nine SWGs span the full breadth of astrophysical topics relevant to a next-generation (sub-)millimeter facility—from the first galaxies all the way to the formation of planetary systems, and including time-domain astrophysics. Descriptions of the different working groups can be found in the Appendix of this document. Community engagement has been substantial: more than 400 researchers—so far primarily from Europe—have joined the network.

In March 2025, more than 80 science pitches were submitted by members of the network to the SWGs, forming the initial basis for identifying the key scientific opportunities of the 2040s. Two major in-person workshops were held in 2025 that synthesized this community input and defined the first scientific priorities for ALMA2040. At a workshop in Heidelberg in May 2025 [3], representatives from the SWGs distilled the >80 science pitches into a preliminary set of unifying science themes. During a workshop at the Lorentz Center in Leiden in November 2025 [4], these themes were refined and validated by the working group leads and external experts, and were linked to realistic technical pathways. This process established the scientific foundation for developing unified science requirements and the ALMA2040 reference design concept. The Scientific Working Groups have delivered a series of 27 White Papers for the ESO Expanding Horizons call (submitted 15 December 2025); the titles of the submitted White Papers can be found in the Appendix of this document; the full abstracts can be found on the ALMA2040 website [1].

# 2. Key Science Objectives for a Next-Generation ALMA-like Facility

As an outcome of the Lorentz Center workshop, it was determined that the four Key Science Objectives (KSOs) as defined during the ALMA2040 workshop in Heidelberg in May 2025 remain compelling for a transformational facility operational in the 2040s. Broadly, these are:

- ***The emergence and evolution of galaxies and black holes***
  From the detailed characterization of galaxies and their environments—including primordial systems out to z ~ 20—to the formation and impact of black holes and magnetic fields, and the kinematics and dark-matter halos of Milky Way progenitors.

- ***The evolution of the cosmic baryon cycle in galactic ecosystems***
  From the origin of the first dust—including pathways for dust production and stellar mass loss—to the multi-phase gas cycles in and around galaxies, and the rise of chemical complexity in the star-forming interstellar medium that ultimately seeds life.

- ***The life cycle of planetary systems***
  From the formation of planetary systems and atmospheres—all the way down to rocky planets and their paleo-Biospheres, and the emergence of complex organics across solar system bodies—to their destruction, including the impact of solar and stellar physics and evolution on the planetary life cycle.

- ***The physics of the extreme universe***
  From probing the birth, acceleration, and feedback of jets across the cosmic mass scale to unveiling the millimeter signatures of explosive transients, shocks, and coherent bursts from neutron stars in the emerging era of wide-field time-domain and multi-messenger astronomy.

These Key Science Objectives will continue to be refined as part of the development and definition of the Science & Systems Requirements before Spring 2026.

## 3. Technical Working Groups

Beginning after the Heidelberg workshop, four Technical Working Groups (TWGs) have been established in order to translate the science objectives into a feasible technical concept. Their role is to define system-level requirements, explore technological solutions, and develop the first reference design. Specifically:

- ***Science & System Requirements Working Group (SSRWG)***
  ***Leads:*** **María Díaz Trigo** & **Martina Wiedner**
  **E-mail: SSRWG+owner@euro-alma2040.groups.io**
  This TWG works closely with all SWGs to define unified science requirements and derive high-level system and performance requirements, including antenna configurations. The SSRWG forms the central interface between science and technical development.

- ***Instrument Working Group (IWG)***
  ***Leads:*** **Andrey Baryshev** & **Peter Huggard**
  **E-mail: iwg+owner@euro-alma2040.groups.io**
  This TWG is responsible for evaluating and defining the instrumentation needed for a next-generation facility, including antennas, receiver systems, correlator design, firmware, and real-time control systems. Cost efficiency in construction and operations is a key consideration.

- ***Operations Working Group (OWG)***
  ***Leads:*** **Elizabeth Humphreys** & **Gergö Popping**
  **E-mail: OWG+owner@euro-alma2040.groups.io**

This TWG explores operational models for the facility, including multi-messenger coordination, sustainability and environmental considerations, maintenance models, and pathways for high operational efficiency.

- ***Data Archiving and Processing Working Group (DAPWG)***
  ***Leads:*** **TBA**
  **E-mail: TBA**
  This TWG defines requirements for off-line processing, calibration, data archiving, data management, and the science archive to ensure a robust and efficient end-to-end data flow. *Any researcher interested in this TWG, in particular its leadership, please contact any member of the steering committee (details in the Appendix).*

## 4. Path Toward a Reference Design Concept

The Scientific and Technical Working Groups are now working toward the first ALMA2040 reference design concept, which will form the core of the proposal to ESO's Expanding Horizons initiative in mid-2027. Over the coming year, these groups will translate the scientific priorities into unified science requirements, system-level specifications, and finally an initial instrument architecture.

By the end of December 2025, an initial set of science requirements will be completed. In early 2026, the Scientific Working Groups will evaluate potential antenna configurations. By the end of the first quarter of 2026, the collaboration aims to deliver a consolidated set of science requirements and a first outline of the reference design concept.

These components will be reviewed at an in-person workshop in spring 2026, marking a key step toward a coherent and community-endorsed base concept. Although the schedule is ambitious, the established milestones and coordinated structure provide a clear pathway toward a complete design concept for the 2027 submission.

## 5. Upcoming in 2026: Dedicated ALMA2040 White Papers

The outcomes of the Scientific and Technical Working Groups will be solidified into a suite of ALMA2040 White Papers that together articulate the scientific justification, technical requirements, and conceptual design for the facility.

### Science White Papers

As mentioned above, 27 White Papers were submitted to the ESO Expanding Horizons call on 15 December 2025 (see the Appendix for titles; abstracts are on the website [1]). These form the starting point and will be expanded during the first half of 2026 into full ALMA2040 Scientific White Papers that will be published on arXiv. New topics and contributions are more than welcome (see below).

A dedicated simulation tool will be released at the beginning of 2026 to support the development of science requirements and technical specifications.

### Technical White Papers

The TWGs will prepare Technical White Papers for arXiv submission in fall 2026, detailing the instrument concept, system requirements, operations model, and data-processing architecture.

### Community Participation

All members of the global mm/sub-mm community are warmly invited to contribute. Please contact the SWG or TWG leads or the steering committee to get involved. Contact details can be found at the end of this document.

## 6. Timeline of Key Upcoming Milestones

- **Early 2026:** Release of simulation tool
- **March 2026:** In-person workshop on first draft of ALMA2040 base concept
- **Spring 2026:** Finalization of unified science requirements and base concept
- **Mid-2026:** ALMA2040 scientific White Papers posted to arXiv
- **Fall 2026:** ALMA2040 technical White Papers posted to arXiv
- **November 2026:** Submit Letter of Intent to ESO Call for Ideas
- **Mid-2027:** Submission of ALMA2040 design concept to ESO's Expanding Horizons initiative

**Links:**
[1] https://www.euroalma2040.com/home
[2] https://next.eso.org/
[3] https://www.euroalma2040.com/workshops/hda-workshop
[4] https://www.euroalma2040.com/workshops/lorentz-workshop

## Appendix

## White Papers submitted to ESO's Expanding Horizons call

The following White Papers were submitted to ESO's call as part of the Expanding Horizons initiative, grouped here under the four Key Science Objectives:

**KSO1: The emergence and evolution of galaxies and black holes**

**[WP1]** *How do the first galaxies assemble? Revealing the interstellar medium at the dawn of galaxy formation*

- **[WP2]** *What drives stellar and black-hole growth through cosmic time? Mapping the physics and chemistry of gas and dust in distant galaxies*
- **[WP3]** *Probing cold gas and dust in quiescent galaxies across cosmic time: What quenches massive galaxies in the early Universe?*
- **[WP4]** *Tracing the emergence of galaxy discs and dark matter halos in the early Universe*
- **[WP5]** *High-z AGN: The role of massive black holes in galaxy evolution*
- **[WP6]** *AGN Feeding and Feedback: Resolving the Torus*

**KSO2: The evolution of the cosmic baryon cycle in galactic ecosystems**

- **[WP7]** *Resolving the Baryon Cycle over Cosmic Time: What regulates star formation in galaxies across cosmic time?*
- **[WP8]** *Tracking the flow of interstellar gas across scales: from galaxies to star clusters*
- **[WP9]** *What drives star formation across local galaxies? Resolving the underlying physics behind star-forming scaling relations*
- **[WP10]** *Molecular atlas of galaxies: High-resolution wideband molecular line inventories of nearby galaxies*
- **[WP11]** *Chemical complexity from molecular clouds to star-forming cores: Understanding its emergence, evolution, local and galactic variations*
- **[WP12]** *Star Formation across the Galaxy: Resolving fragmentation, accretion and feedback throughout the entire Milky Way*
- **[WP13]** *Evolved Stars: Dust formation/mass loss*

**KSO3: The life cycle of planetary systems**

- **[WP14]** *Planet formation on solar system scales: How do gas giants form and shape their planetary system architectures?*
- **[WP15]** *Probing the chemistry of planet-forming material on solar system scales: What is the elemental and isotopic composition of terrestrial planet-building material?*
- **[WP16]** *Planet forming disk populations and their connection to the environment: How do planets form in the most common star-forming conditions in the galaxy?*
- **[WP17]** *Planetary system architectures in the final stages of planet formation: How do late-stage processes shape planet assembly and volatile delivery?*
- **[WP18]** *The complex dynamics of the early stages of star and planet formation*
- **[WP19]** *Evolved stars, multiplicity and planets: Shaping and Being Shaped by Stellar Evolution*
- **[WP20]** *From climate to chemistry: understanding atmospheric variability, dynamics, and volatile and organics origins across worlds*
- **[WP21]** *The energetic atmosphere of the Sun: How do small-scale dynamic processes heat and structure the solar atmosphere?*

**KSO4: The physics of the extreme universe**

- **[WP22]** *Shaping the future of Global Interferometric Arrays: Imaging Strong Gravity and Magnetic Fields*
- **[WP23]** *Strong gravitational lensing with a next-generation sub-millimetre/millimetre array: Astrophysical tests of the nature of dark matter in the 2040s*
- **[WP24]** *The Magnetized Universe: From planet-forming disks to the high-redshift universe*
- **[WP25]** *Pulsars & Fast transients: Understanding fundamental physics using mm-wavelength observations of neutron stars*
- **[WP26]** *Understanding cosmic explosions*
- **[WP27]** *mm-jets in the multi-messenger era: The launching, collimation, acceleration, feedback and evolution of astrophysical jets across all scales*

## Scientific Working Groups

The SWGs unite experts focused on particular science topics, aiming to determine the key science goals motivating a next-generation millimeter/submillimeter array, the technical specifications required, and any adaptations resulting from evolving design considerations. The SWGs, which naturally include overlap and scientific synergies, are:

- ***The High-Redshift Universe WG***
  ***Leads:*** **Tom Bakx** & **Francesca Rizzo**
  **E-mail: high-z+owner@euro-alma2040.groups.io**
  Studies galaxies, supermassive black holes, and large-scale structure from the earliest observable epochs ($z > 15$) to the nearby universe ($z \sim 0.3$).

- ***Active Galactic Nuclei WG***
  ***Leads:*** **Roberto Decarli** & **Miguel Pereira Santaella**
  **E-mail: agn+owner@euro-alma2040.groups.io**
  Explores the structure, environment, and evolution of AGN across cosmic time, and their impact on host galaxies.

- ***Cosmology and Fundamental Physics WG***
  ***Leads:*** **Violette Impellizzeri** & **Hannah Stacey**
  **E-mail: cosmology-fundamental-physics+owner@euro-alma2040.groups.io**
  Investigates topics such as fundamental constants and the nature of dark matter; probes such as gravitational lensing; EHT applications.

- ***The Local Universe WG***
  ***Leads:*** **Jan Forbrich** & **Miguel Querejeta**
  **E-mail: local-universe+owner@euro-alma2040.groups.io**

Examines the structure, kinematics, and environment of galaxies in the Local Group and beyond ($z < 0.3$), bridging detailed Milky Way studies with observations of distant galaxies.

- ***Interstellar Medium and Star Formation WG***
  ***Leads:*** **Maite Beltran** & **Jes Jørgensen**
  **E-mail:** **ism-sf+owner@euro-alma2040.groups.io**
  Probes the physics and chemistry of the Galactic ISM to understand star formation, from circumstellar disks to chemical complexity across environments.

- ***Planet Formation WG***
  ***Leads:*** **Luca Matrà** & **Catherine Walsh**
  **E-mail:** **planet-formation+owner@euro-alma2040.groups.io**
  Studies the formation and evolution of planetary systems, from protostellar disks to mature planetesimal belts, using continuum and molecular line observations to trace dust, gas, and dynamics.

- ***Sun and Stars WG***
  ***Leads:*** **Wouter Vlemmings/Theo Khouri** & **Sven Wedemeyer**
  **E-mail:** **sun-stars+owner@euro-alma2040.groups.io**
  Uses solar and stellar observations to study atmospheres, magnetic activity, and late-stage evolution, including processes that drive enrichment and dust production.

- ***Transients and Time-Domain Astronomy WG***
  ***Leads:*** **Karri Koljonen** & **Kuo Liu**
  **E-mail:** **time+owner@euro-alma2040.groups.io**
  Explores the sub-mm regime promise for time-domain studies of pulsars, neutron stars, black holes, FRBs, GRBs, supernovae, and other fast transients.

- ***Solar System Bodies and Exoplanets WG***
  ***Leads:*** **Arianna Piccialli** & **Miriam Rengel**
  **E-mail:** **solarsystem-exoplanets+owner@euro-alma2040.groups.io**
  Investigates planetary atmospheres, moons, asteroids, comets, and exoplanets, focusing on formation, evolution, and comparative planetology.

## Steering Committee

**Stefano Facchini** (stefano.facchini@unimi.it)
**Jacqueline Hodge** (hodge@strw.leidenuniv.nl)
**Eva Schinnerer** (schinner@mpia.de)
**Gie Han Tan** (g.h.tan@tue.nl)